\documentclass[preprint]{aastex631}

\usepackage{xspace}
\usepackage{amssymb}

\newcommand{\co}{CO$_2$\xspace}
\newcommand{\pn}{pN$_2$\xspace}
\newcommand{\wm}{W/m$^2$\xspace}
\newcommand{\water}{H$_2$O\xspace}

\begin{document}

\title{Climate Transition to Temperate Nightside at High Atmosphere Mass}

\correspondingauthor{Evelyn Macdonald}
\email{evelyn.macdonald@mail.utoronto.ca}

\author[0000-0001-5540-3817]{Evelyn Macdonald}
\affiliation{Department of Physics, University of Toronto, Toronto, ON, Canada M5S 1A7}

\author{Kristen Menou}
\affiliation{Department of Physical and Environmental Sciences, University of Toronto, Scarborough, ON, Canada M1C 1A4}
\affiliation{David A. Dunlap Department of Astronomy and Astrophysics, University of Toronto, ON, Canada ON M5S 3H4}
\affiliation{Department of Physics, University of Toronto, Toronto, ON, Canada M5S 1A7}

\author[0000-0003-0029-5278]{Christopher Lee}
\affiliation{Department of Physics, University of Toronto, Toronto, ON, Canada M5S 1A7}

\author[0000-0001-6774-7430]{Adiv Paradise}
\affiliation{David A. Dunlap Department of Astronomy and Astrophysics, University of Toronto, ON, Canada ON M5S 3H4}

\begin{abstract}

Our recent work shows how M-Earth climates and transmission spectra depend on the amount of ice-free ocean on the planet's dayside and the mass of N$_2$ in its atmosphere. M-Earths with more ice-free ocean and thicker atmospheres are hotter and more humid, and have larger water vapour features in their transmission spectra. In this paper, we describe a climate transition in high-\pn simulations from the traditional ``eyeball" M-Earth climate, in which only the substellar region is temperate, to a ``temperate nightside" regime in which both the dayside and the nightside are entirely ice-free. Between these two states, there is a ``transition" regime with partial nightside ice cover. We use 3D climate simulations to describe the climate transition from frozen to deglaciated nightsides. We attribute this transition to increased advection and heat transport by water vapour in thicker atmospheres. We find that the nightside transitions smoothly back and forth between frozen and ice-free when the instellation or pCO$_2$ is perturbed, with no hysteresis. We also find an analogous transition in colder planets: those with thin atmospheres can have a dayside hotspot when the instellation is low, whereas those with more massive atmospheres are more likely to be in the ``snowball" regime, featuring a completely frozen dayside, due to the increased advection of heat away from the substellar point. We show how both of these climate transitions are sensitive to instellation, land cover, and atmosphere mass. We generate synthetic transmission spectra and phase curves for the range of climates in our simulations.

\end{abstract}


\section{Introduction} \label{sec:intro}

Recent studies have shown that M-Earths can have ice-free nightsides when the incident flux or greenhouse gas abundance is sufficient \citep{Turbet2016, Komacek2019, Zhang2020, Paradise2022}. Our recent work \citep{Macdonald2022, Macdonald2024} has shown that the combined effects of unconstrained land cover and atmosphere mass result in a large range of possible M-Earth climate states, which will be difficult to differentiate in transit spectra. Following up on these results, this paper focuses on a climate transition to a new climate regime at high atmosphere mass (pN$_2$), featuring a fully deglaciated nightside, moist atmosphere, and small day-to-night temperature gradient. This new regime is a departure from the frozen nightside and partially deglaciated dayside typical of M-Earth climates. We also present a transition regime featuring a partially deglaciated nightside. This climate transition occurs because more massive atmospheres are conducive to increased day-night heat transport through dayside evaporation and advection of water vapour to the nightside. This paper describes these transition-related mechanisms in detail.

Although some M-Earth climates with ice-free or partly ice-covered nightsides have been identified in previous work, and the effects of atmosphere mass have been investigated in asynchronously rotating planets with Sun-like hosts, as discussed below, we are unaware of any detailed description of this climate transition on synchronously rotating planets with M-dwarf hosts to date. In this paper, we characterize the transition for M-Earths and discuss the dependence of the transition threshold on dayside land cover. We describe the physical effects of \pn on climate in Section \ref{sec:climateeffects}, describe our simulations and climate regimes in Section \ref{sec:simregime}, discuss climate transition mechanisms and validation tests in Section \ref{sec:validation}, present synthetic observations in Section \ref{sec:obs}, and discuss our results in Section \ref{sec:discuss}.

\section{Climate Effects of N$_2$ Partial Pressure}\label{sec:climateeffects}

N$_2$ is not a greenhouse gas, and it is difficult to detect in transmission spectra due to a lack of strong infrared absorption features \citep{Benneke2012}. It can nonetheless have significant climate effects when it makes up a large proportion of the atmosphere, as it does on Earth. As a background gas, N$_2$ affects dynamics, heat transport, humidity, and surface temperature. N$_2$ has a direct radiative effect through Rayleigh scattering and an indirect radiative effect through pressure broadening of other species. These climate effects are described below.

\subsection{Pressure Broadening}

Pressure broadening of greenhouse gases is a warming effect: interactions with background gas molecules causes a broadening of a gas' absorption lines, resulting in an amplification of the greenhouse effect as more radiation is absorbed by the atmosphere \citep{Goldblatt2009}. \citet{Kopparapu2014} found that in warm wet atmospheres, the pressure broadening of \water vapour lines warms the surface by significantly reducing outgoing longwave radiation. \co lines are also broadened, resulting in an enhanced \co greenhouse effect. The impact on \water is larger because of a positive temperature-water vapour feedback: although the saturation vapour pressure of \water does not depend directly on the background gas pressure, it has a strong dependence on temperature, which itself is very sensitive to \pn. \water abundance is therefore highly sensitive to \pn. More massive atmospheres are warmer and can therefore hold more water vapour, whose greenhouse effect is then amplified by pressure broadening \citep{Keles2018}.

\subsection{Rayleigh Scattering}

Rayleigh scattering is a cooling effect with a $\lambda^{-4}$ dependence, meaning that shortwave radiation is preferentially scattered, regardless of atmospheric composition. A higher surface pressure means that more molecules are available to scatter photons; consequently, the cooling effect increases with the partial pressure of N$_2$ or other gases. The cooling effect of Rayleigh scattering at high \pn dominates over the warming effect of pressure broadening for planets with Sun-like spectra \citep{Keles2018, Komacek2019, Paradise2021}. However, since Rayleigh scattering is weaker for M-Earths due to the redder spectra of M-dwarfs \citep{Ramirez2020}, the warming effect of pressure broadening can be expected to continue to outweigh the cooling effect of Rayleigh scattering at high surface pressures for M-Earths.

\subsection{General Circulation}

There is a lapse rate warming feedback associated with increased \pn: the moist adiabatic lapse rate increases with pressure, bringing it closer to the dry adiabatic lapse rate. On Earth, this results in decreased convection in the tropics and therefore a warmer surface \citep{Goldblatt2009, Xiong2022}. Meanwhile, the atmosphere's total heat capacity increases with mass, leading to a decrease of high-latitude lower-troposphere radiative cooling \citep{Chemke2016}.

\citet{Zhang2020} describe a change in the general circulation of a synchronously rotating M-Earth with increasing \pn: the substellar upwelling region becomes larger, the downwelling region becomes smaller, and the horizontal and vertical winds become much slower. These changes result in a higher upper-troposphere relative humidity, and therefore in warming due to an enhanced water vapour greenhouse effect. They also note that despite the reduction in wind speed, the magnitude of the mass streamfunction increases with \pn because the atmosphere is more massive; consequently, the day-night temperature gradient is reduced because more heat is transported to the nightside. This circulation trend is analogous to the trend for Earth-like planets, on which thicker atmospheres lead to slower zonal winds, greater mass transfer, more efficient heat transport, and smaller equator-to-pole temperature gradients \citep{Kaspi2015, Wordsworth2016}.

\section{Description of Simulations}\label{sec:simregime}

\subsection{Simulation Parameters}

We use the 3D general circulation model (GCM) ExoPlaSim \citep{Paradise2022}. The parameters for our simulations are described in Table \ref{tab:pn2simulations}. We use the Group C simulations from \citet{Macdonald2024} and variations on these. The 0.2~M$_\oplus$ planet is synchronously rotating around a 2600~K star. This system is optimized for transit spectroscopy: the planet's transit depth is largest with a small host star, and the planet's low mass results in a large atmospheric scale height, which makes spectral features easier to detect than on an Earth-sized planet.

We vary the dayside land fraction from 0 to 100\% using the Substellar Continent (SubCont) and Substellar Ocean (SubOcean) configurations described in Figure 1 of \citet{Macdonald2022}. SubCont is a circular continent centred at the substellar point with ocean everywhere else. SubOcean is the opposite: a circular ocean centred at the substellar point with land everywhere else. The original Group C simulations all receive 881.7~W/m$^2$ of instellation to match Proxima Centauri b, with the orbital period and rotation rate adjusted for a 2600~K star. In this paper, we also include variations with instellations of 700, 800 and 1000~W/m$^2$, with the rotation periods adjusted accordingly and all other parameters unchanged, to explore how colder and warmer climates respond to \pn variations. 

\begin{table}[h!]
\centering
\begin{tabular}{c||c|c|c}
\centering
Parameter & Group C \citep{Macdonald2024} & Variations & Perturbations \\
\hline
\hline
Radius (R$_\oplus$) & 0.646 & -- & -- \\
Mass (M$_\oplus$) & 0.2 & -- & -- \\
Gravity (m/s$^2$) & 4.7 & -- & --  \\
Period (days) & 4.96 & 5.89, 5.33, 4.51 & --\\
pN$_2$ (bar) & 0.5, 1, 2, 4, 6, 8 & -- & --\\
pCO$_2$ (millibar) & 1 & -- & 0.2 -- 3 \\
Stellar temperature (K) & 2600 & -- & -- \\
Stellar radius (R$_\odot$) & 0.1 & -- & -- \\
Stellar flux (W/m$^2$) & 881.7 & 700, 800, 1000 & 810 -- 910 \\
Resolution (lat$\times$lon) & $64\times128$ & -- & $32\times64$
\end{tabular}
\caption{Description of simulations.}
\label{tab:pn2simulations}
\end{table}

We also include some lower-resolution simulations to explore cycles between climate states, by varying either the pCO$_2$ or the instellation. Although the quantitative threshold for the climate transition can be resolution-dependent, the mechanism of the transition is the same at both resolutions. Our results are therefore qualitatively valid, and the more efficient configuration makes multi-millennium simulations feasible.

\subsection{Climate Regimes}

Our simulations generally fall into four distinct climate regimes (Figure \ref{fig:climate_regimes}). Each planet's climate state is determined by the combined effects of land cover, instellation, \pn, and p\co. These regimes are as follows:

\begin{figure}[h!]
    \includegraphics[width=\textwidth]{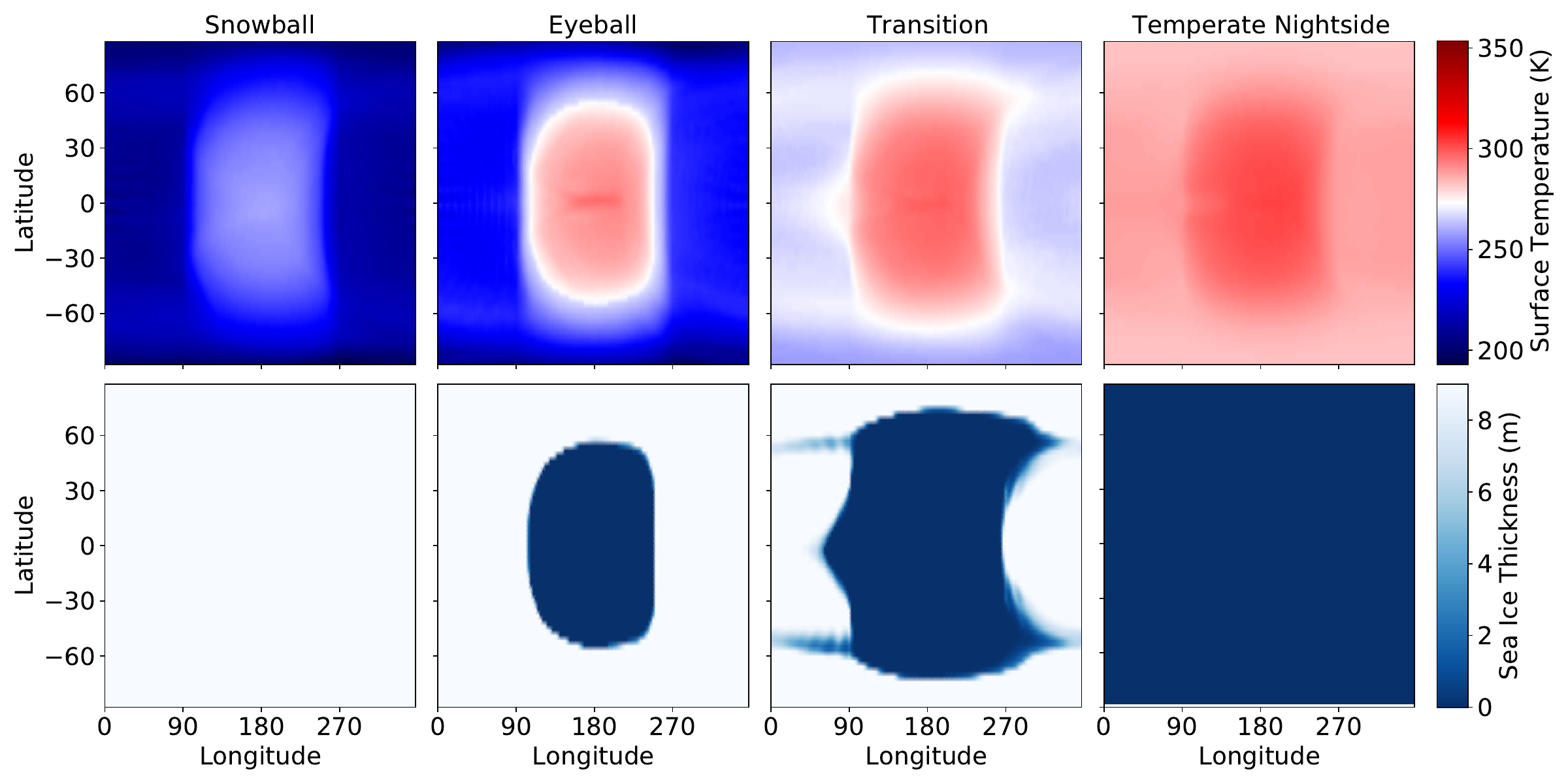} 
    \caption{Surface temperature (top row) and sea ice (bottom row) maps of aquaplanets in the snowball, eyeball, transition, and temperate nightside climate regimes.}
    \label{fig:climate_regimes}
\end{figure}

\begin{itemize}
    \item The ``eyeball" regime \citep{Pierrehumbert2011}, featuring a frozen nightside and a partially deglaciated dayside, is the basic climate regime expected for M-Earths, given the distribution of incident flux on synchronously rotating planets. The simulations in \citet{Macdonald2022, Macdonald2024} are mostly in this regime. The size of the deglaciated region depends on the instellation, \pn, and amount of ice-free ocean on the planet's dayside. Eyeball climates with substellar continents have drier atmospheres and larger day-night temperature contrasts than those with large substellar oceans.
    
    \item The ``snowball" regime features below-freezing daysides and nightsides. This regime is found in low-instellation simulations, especially those with high pN$_2$.
    
    \item The ``temperate nightside" regime describes models with above-freezing temperatures everywhere on the surface. This regime is found in high-instellation, high-pN$_2$ models.

    \item The ``transition" regime features a deglaciated dayside and a partially deglaciated nightside. The frozen regions of the nightside are near the poles and the antistellar point; they are not radially symmetric because of the influence of atmospheric circulation patterns. This regime spans a smaller parameter space than the others.
\end{itemize}

\begin{figure}[h!]
    \centering
    \includegraphics[width=\textwidth]{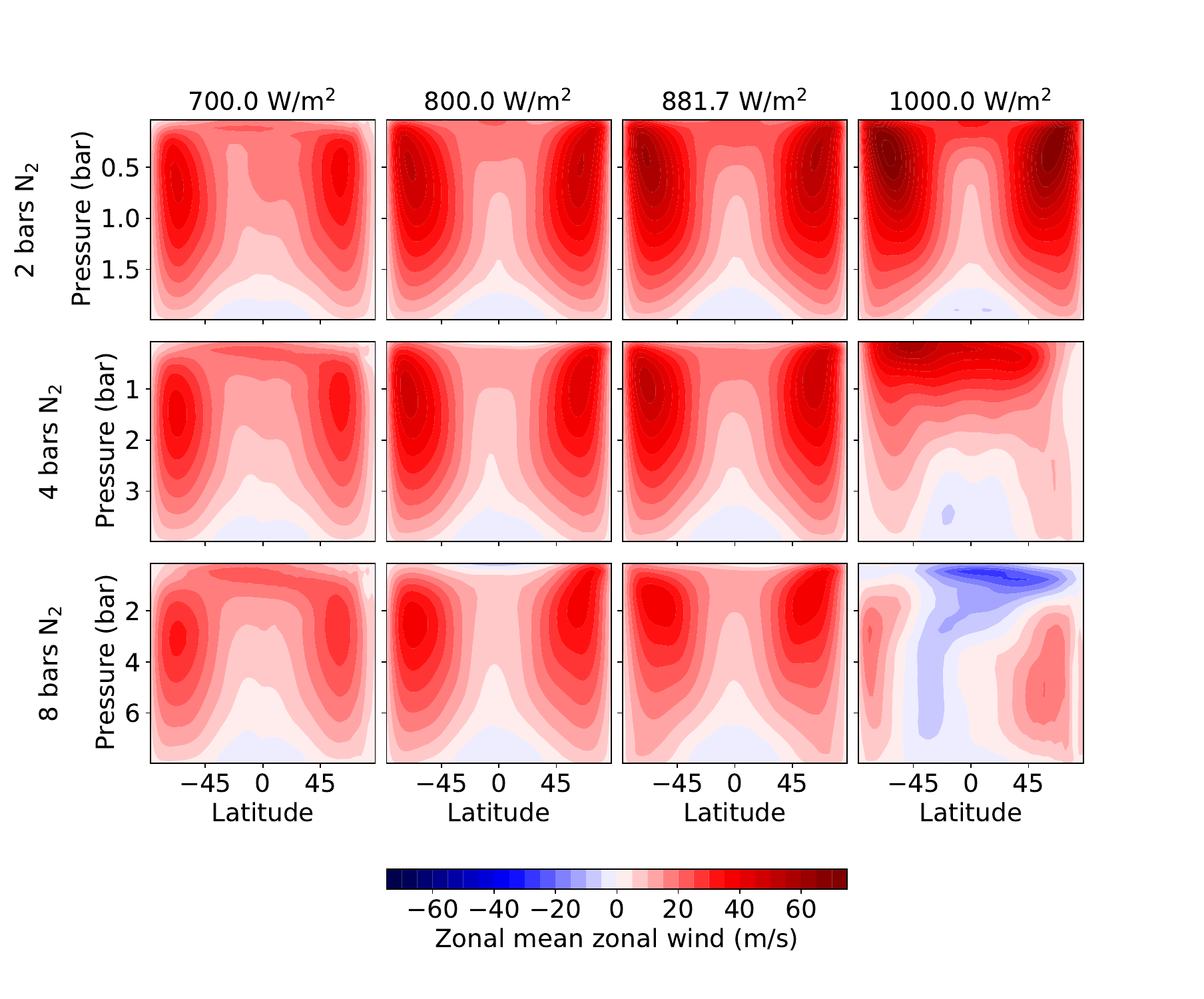}
    \caption{Zonal mean zonal wind for aquaplanets with varying pN$_2$ (rows) and instellation (columns). Most have the Rhines rotator circulation pattern of two jet streams. The jets become slower with increasing \pn and faster with increasing flux, until the circulation breaks down at high flux and \pn due to the transition to the temperate nightside regime.}
    \label{fig:circulation_zmz}
\end{figure}

\begin{figure}[h!]
    \centering
    \includegraphics[width=\textwidth]{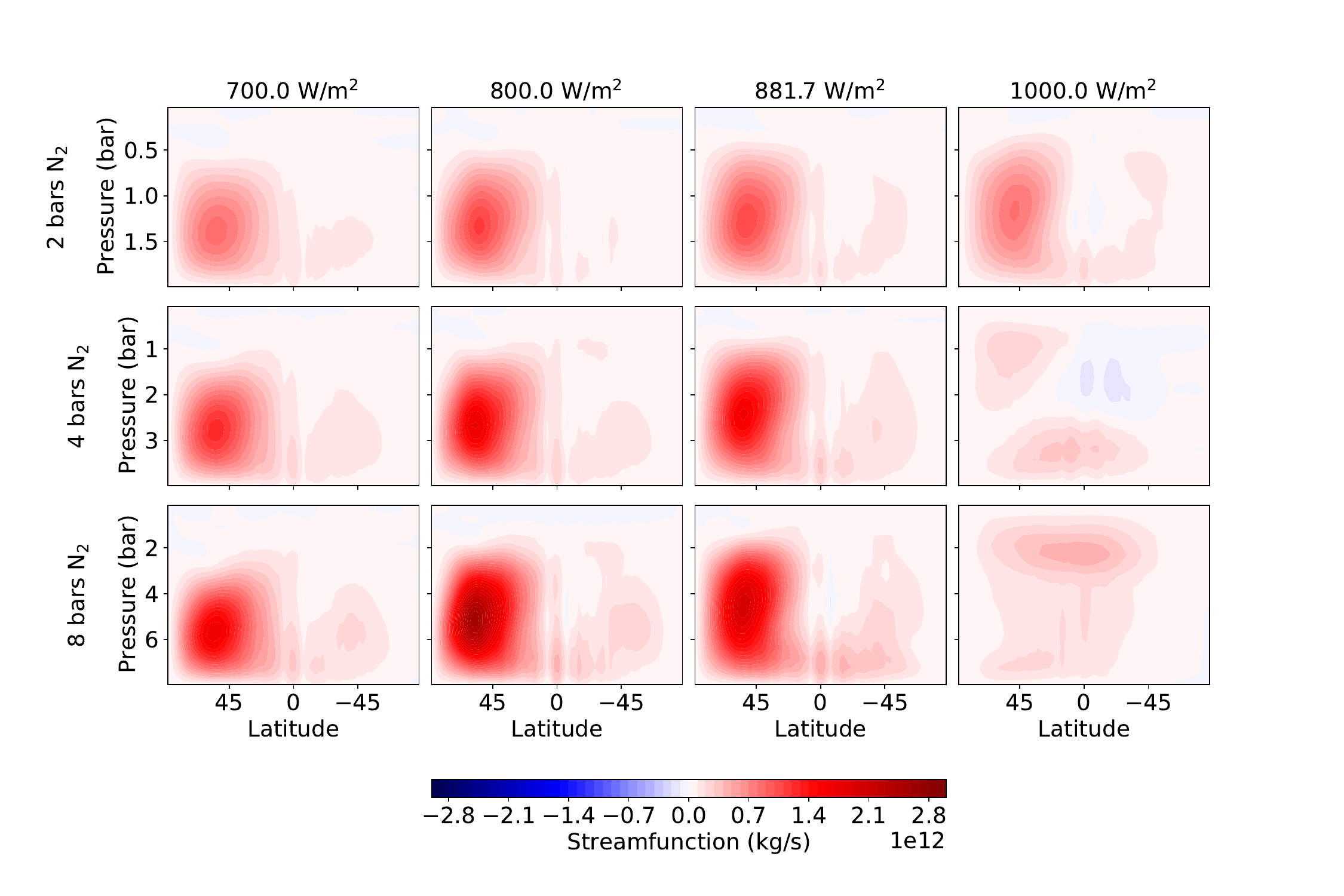}
    \caption{Tidally locked streamfunction \citep{Hammond2021, Paradise2022} for aquaplanets with varying pN$_2$ (rows) and instellation (columns). Most have the Rhines rotator circulation pattern of an overturning cell from the substellar point toward the nightside. The circulation intensifies with both increasing flux and increasing \pn, even though the latter results in slower winds, because the amount of mass being circulated is increasing. There is a sharp transition to a much weaker streamfunction associated with the transition to the temperate nightside regime (bottom right).}
    \label{fig:circulation_tlsf}
\end{figure}

Figures \ref{fig:circulation_zmz} and \ref{fig:circulation_tlsf} show the circulation patterns of aquaplanets with a range of instellations and \pn. We obtain the trend of reduced wind speeds and increasing mass transport streamfunction magnitude pointed out by \citet{Zhang2020}. As in the case of Earth-like planets \citep{Kaspi2015, Komacek2019}, the strength of the circulation and the zonal wind speeds mostly increase with increasing incident flux; however, we find that the circulation breaks down in the temperate nightside regime at high flux and \pn.

\begin{figure}[h!]
    \centering
    \includegraphics[width=\textwidth]{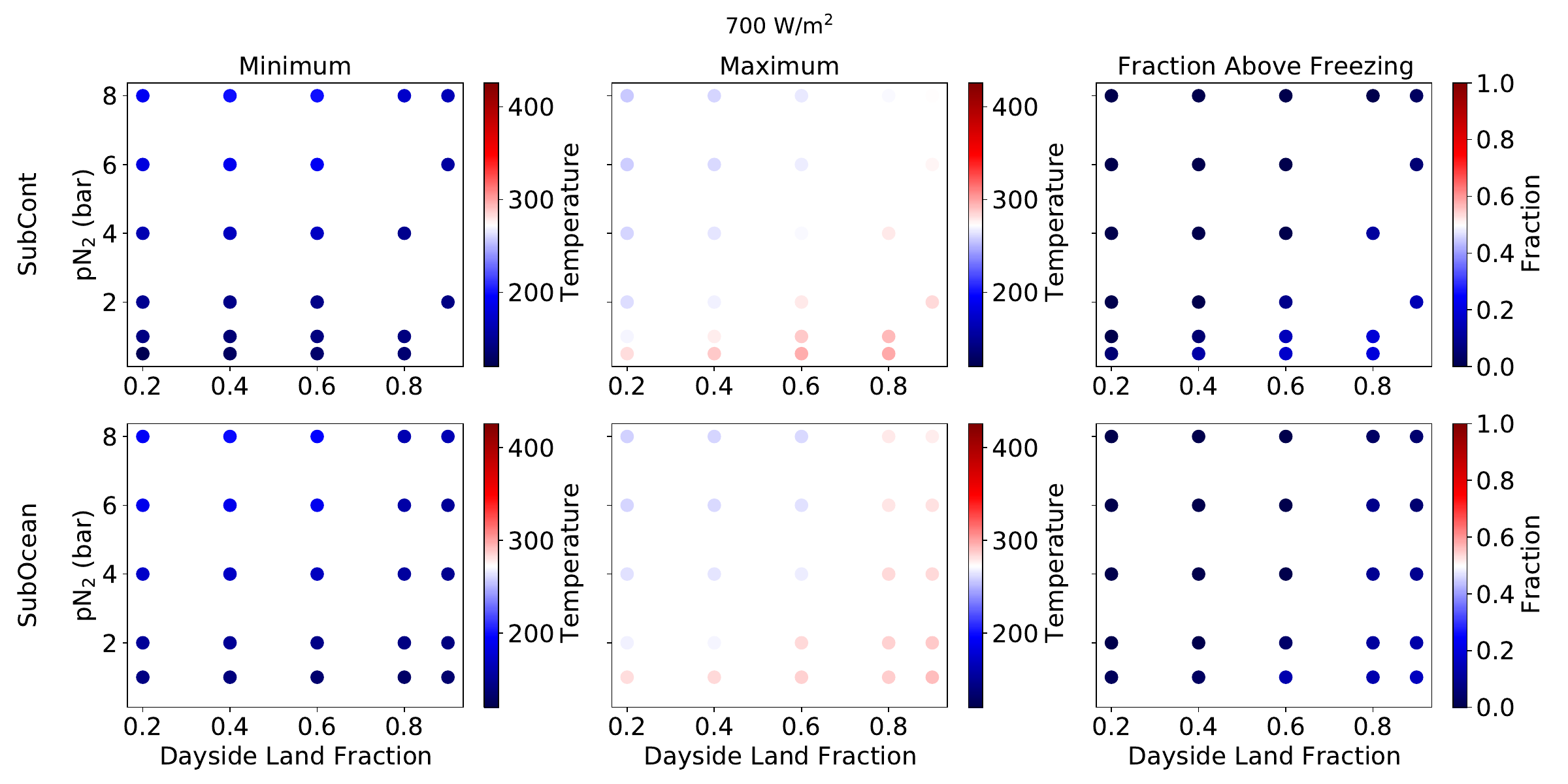}
    \caption{Left to right: minimum and maximum surface temperature and fraction above freezing for SubCont (top row) and SubOcean (bottom row) simulations with an incident flux of 700 \wm. The low-\pn models have the largest surface temperature ranges. Several high-\pn models of both landmap classes have maximum surface temperatures below freezing, meaning that they are in a snowball state.}
    \label{fig:minmax700}
\end{figure}

\begin{figure}[h!]
    \centering
    \includegraphics[width=\textwidth]{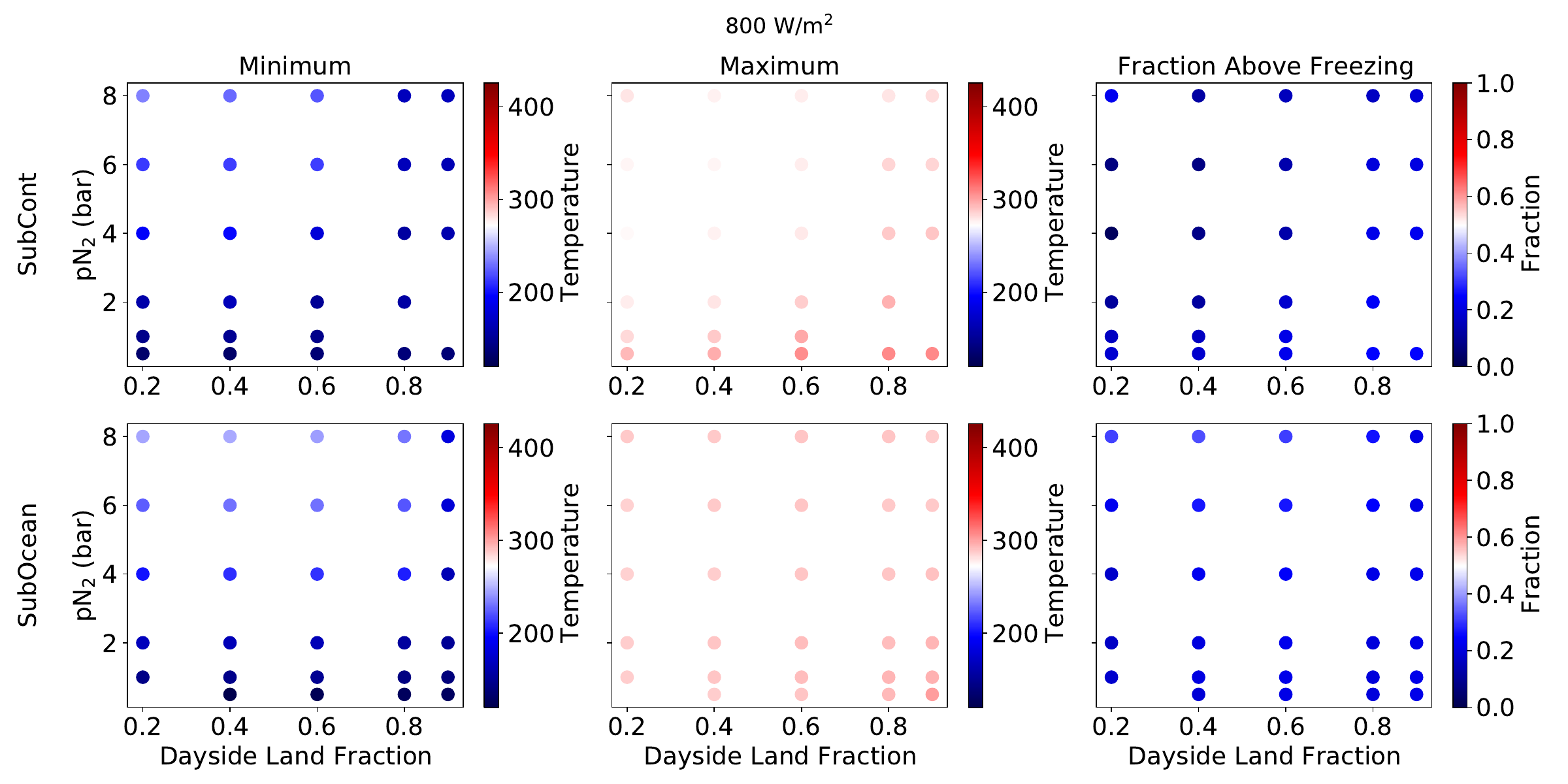}
    \caption{Left to right: minimum and maximum surface temperature and fraction above freezing for SubCont (top row) and SubOcean (bottom row) simulations with an incident flux of 800 \wm. The low-\pn models again have the largest temperature ranges. In particular, SubCont models have both lower minimum and higher maximum surface temperatures at low \pn due to the reduced heat transport of the thinner atmospheres. This effect is less pronounced on SubOcean models due to the presence of dayside ocean.}
    \label{fig:minmax800}
\end{figure}

\begin{figure}[h!]
    \centering
    \includegraphics[width=\textwidth]{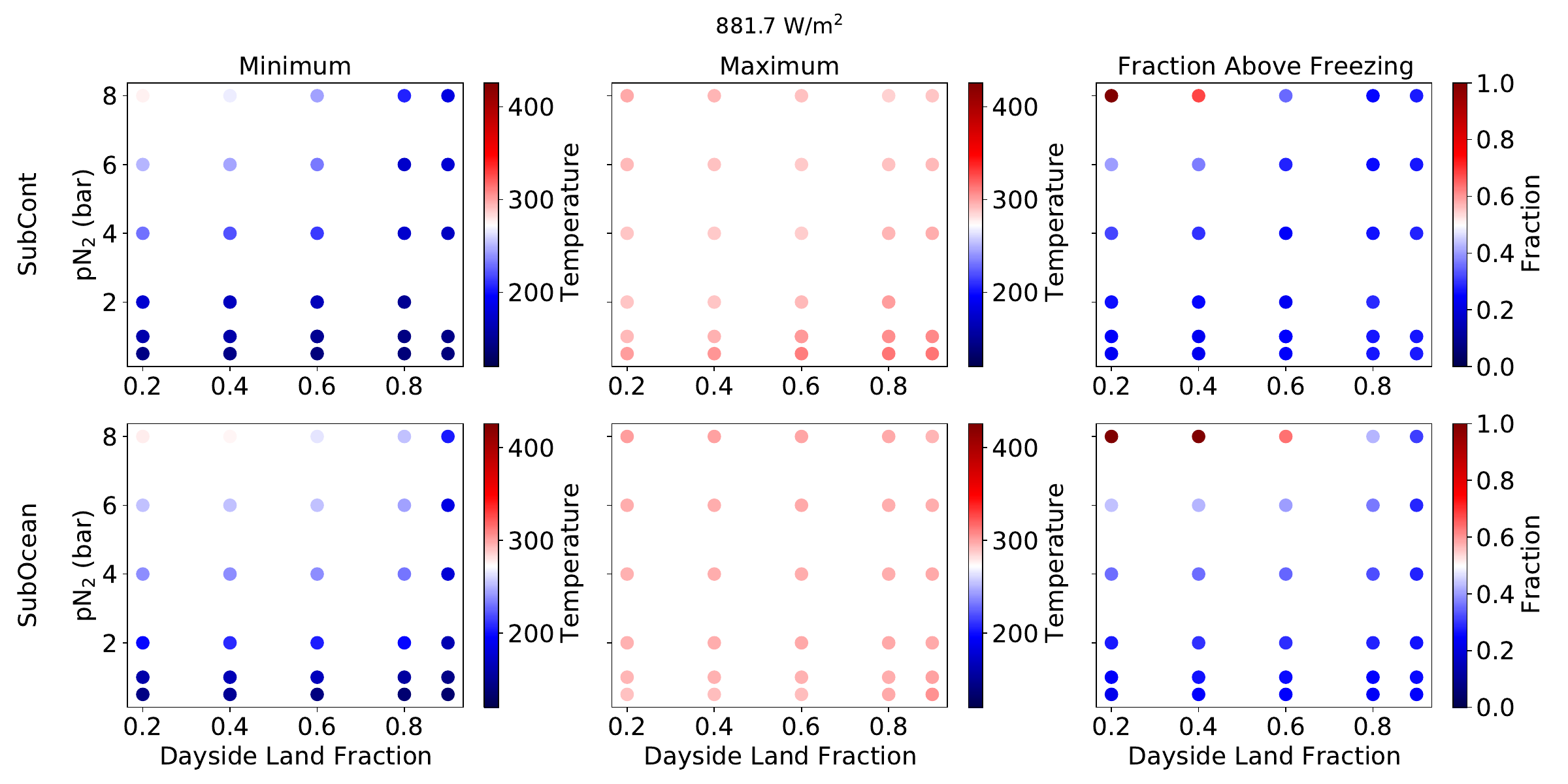}
    \caption{Left to right: minimum and maximum surface temperature and fraction above freezing for SubCont (top row) and SubOcean (bottom row) simulations with an incident flux of 881.7 \wm. Only the highest-\pn, lowest-land-fraction models are in the temperate nightside regime. These have minimum surface temperatures above freezing and substantially reduced day-night temperature contrasts.}
    \label{fig:minmax881.7}
\end{figure}

\begin{figure}[h!]
    \centering
    \includegraphics[width=\textwidth]{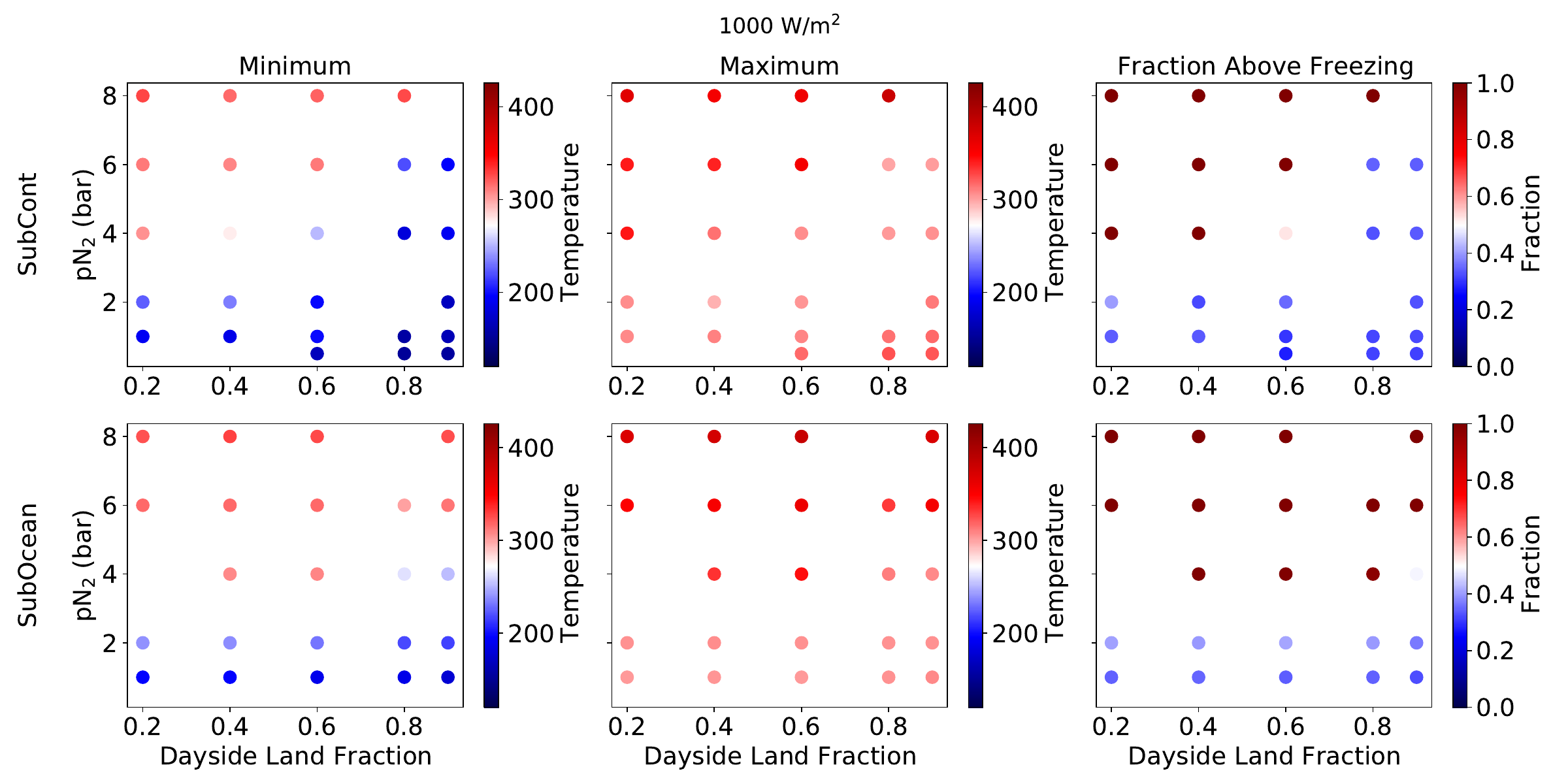}
    \caption{Left to right: minimum and maximum surface temperature and fraction above freezing for SubCont (top row) and SubOcean (bottom row) simulations with an incident flux of 1000 \wm. The high-\pn models are in the temperate nightside regime, with the entire planet above freezing. The threshold for the transition to this regime is at higher \pn for high-land-fraction SubCont models because their atmospheres are drier.}
    \label{fig:minmax1000}
\end{figure}

Figures \ref{fig:minmax700}, \ref{fig:minmax800}, \ref{fig:minmax881.7}, and \ref{fig:minmax1000} show the minimum and maximum surface temperature, and the fraction of the planet above the freezing point of water, for SubCont and SubOcean simulations with instellations of 800, 881.7, 900, and 1000~\wm, respectively. We use fraction above freezing rather than sea ice cover for consistency between planets with different landmaps; since SubOcean planets have nightside land, they cannot have nightside sea ice even when their nightsides are below freezing.

At 700~\wm, simulations are either snowballs (maximum surface temperature below freezing) or eyeballs (frozen nightside with a region above freezing on the dayside). High-\pn and low-land-fraction planets are in a snowball state, since these conditions allow more heat redistribution to the nightside. Low-\pn and high-land-fraction planets of both landmap types are in an eyeball state, with colder nightsides, warmer daysides, and a deglaciated substellar region. This trend suggests that the outer edge of the habitable zone is at lower instellations for high-land-fraction planets with thin atmospheres than for aquaplanets or those with high \pn, since reduced heat transport in the former case keeps the substellar region habitable.

All 800~W/m$^2$ simulations are in the eyeball regime. These planets are relatively cold and dry. Their deglaciated substellar regions vary in size. As in previous studies, some simulations have dayside ice-free ocean, but high-land-fraction SubCont models do not. As with the 700~\wm simulations, maximum surface temperature decreases as \pn increases, such that the substellar region is barely above freezing for low-land-fraction, high-\pn planets due to the increased nightward heat transport of their thicker atmospheres.

Most of the 881.7~\wm simulations are eyeball climates, with the exception of some with low land fraction and high \pn, which are in the transition or temperate nightside regime. The nightsides are cold at low \pn and high land fraction, especially for SubCont simulations, due to their drier atmospheres. 

At 1000~\wm, many more simulations are in the temperate nightside or transition regime,. As before, high \pn and low dayside land fraction contribute to nightward heat transport. The dependence of the threshold \pn on land cover is highlighted for these hotter climates. SubOcean simulations transition to the temperate nightside regime more easily than SubCont simulations because of the difference in land position between these landmap classes. However, at this high instellation, all simulations with a \pn of 8 bars are in the temperate nightside regime regardless of land fraction. 

\section{Transition Validation}\label{sec:validation}

The eyeball and snowball regimes have been studied extensively (e.g., \citealt{Pierrehumbert2011, Checlair2017, Checlair2019, Komacek2019, Yang2019, Sergeev2022, Turbet2022}). The transition and temperate nightside regimes have been seen in other work \citep{Paradise2022, Haqq-Misra2022SAMOSA}, but have not been studied in as much detail. This section uses tests of atmospheric drag, water vapour, radiation balance, and cycling between climate states to isolate mechanisms involved in the transition.

\subsection{Drag}\label{sec:drag}

\begin{figure}[h!]
    \centering
    \includegraphics[width=\textwidth]{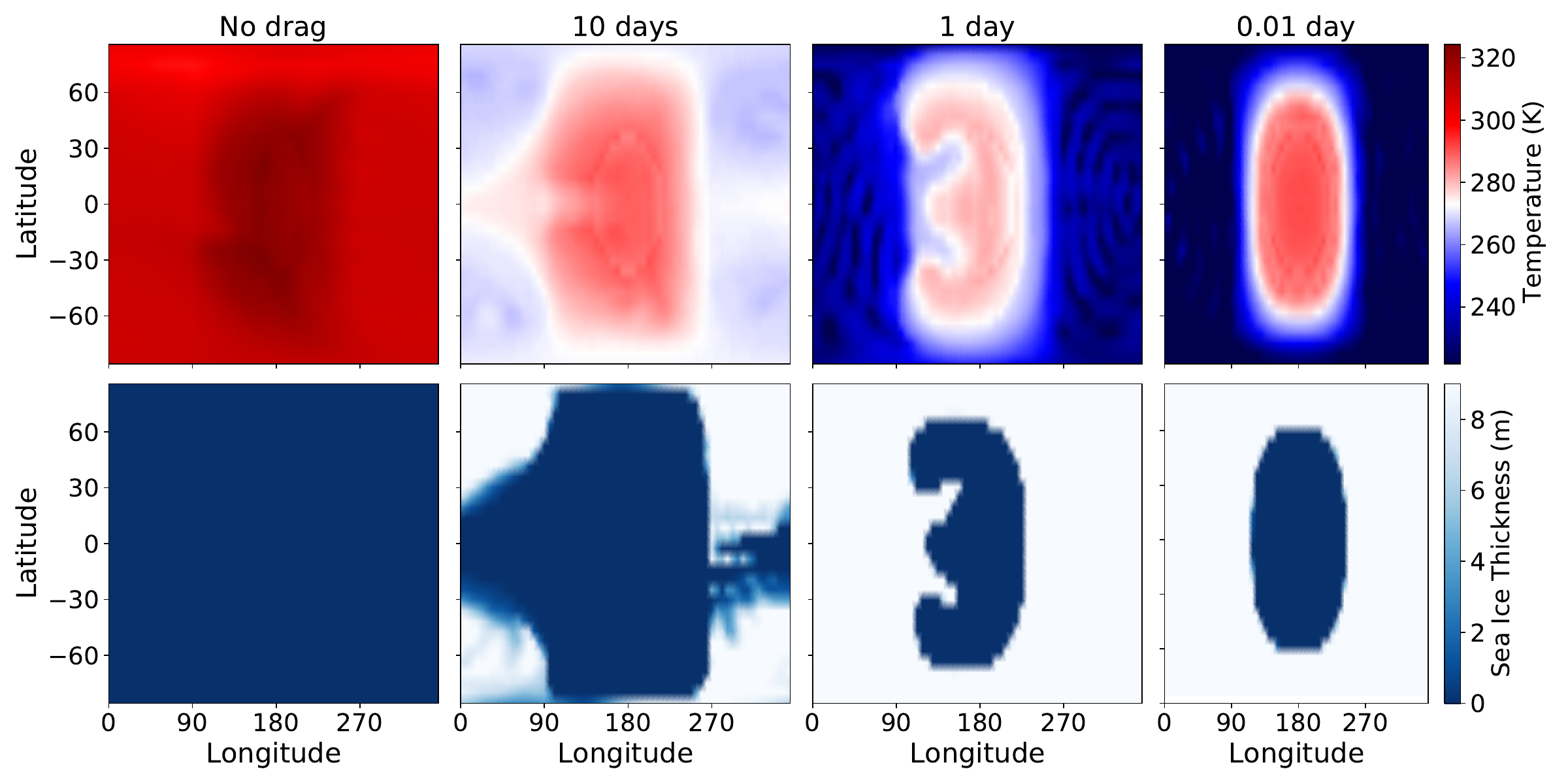}  
    \caption{Surface temperature and sea ice thickness for simulations with varying drag timescales (column titles), for a SubCont planet with 20\% dayside land, 8 bars of N$_2$, and an instellation of 881.7~\wm. As the drag timescale decreases, heat transport to the nightside is inhibited and the planet remains in an eyeball state, which suggests that advection plays an important role in the climate transition.}
    \label{fig:drag}
\end{figure}

\begin{figure}[h!]
    \centering
    \includegraphics[width=\textwidth]{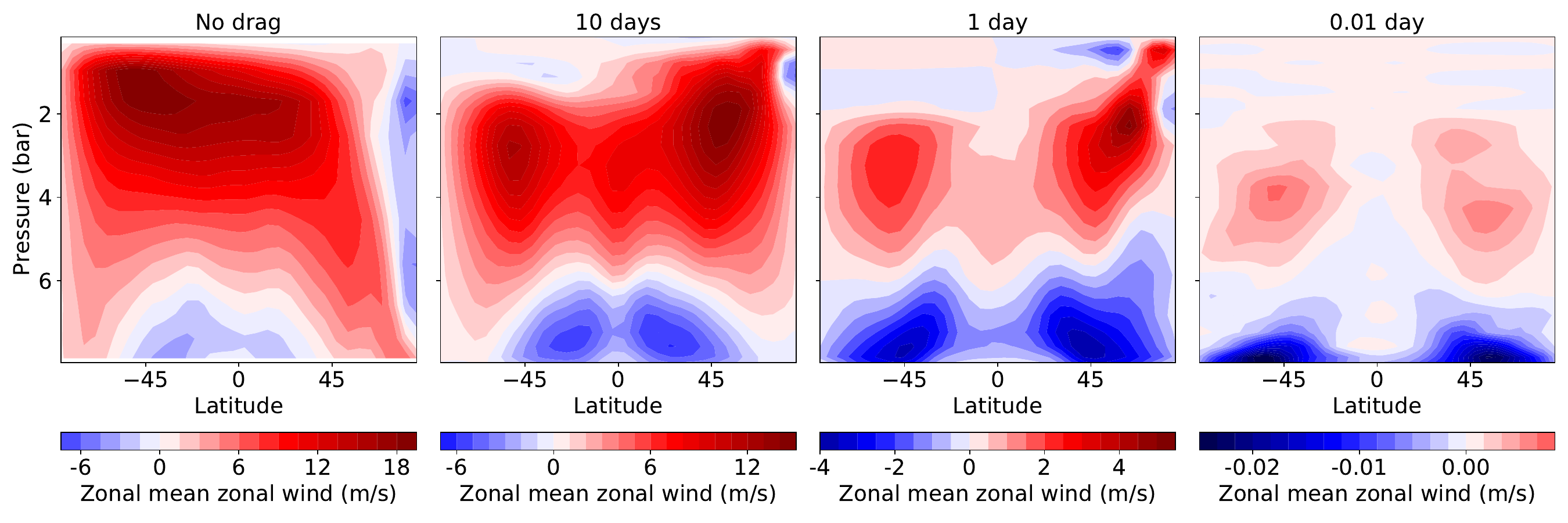}
    \caption{Zonal mean zonal wind for a SubCont planet with 20\% dayside land, 8 bars of N$_2$, and an instellation of 881.7~\wm. Wind speeds are greatly reduced when a drag is applied (column titles), leading to decreased advection of heat to the nightside.}
    \label{fig:drag_zmz}
\end{figure}

We identify advection as a driver of the climate transition by imposing an artificial drag to inhibit circulation. This term, which is not included in the default configuration, causes the horizontal winds to decay with a specified timescale $\tau_{drag}$ at all vertical levels. Our control simulation is a SubCont planet with with 20\% dayside land cover, 8 bars of N$_2$, and an instellation of 881.7~\wm, run at a resolution of T21. Without artificial drag, this planet is in the temperate nightside climate regime. We run three new versions of this simulation with different values of $\tau_{drag}$ and find that the zonal mean zonal wind speeds decrease with decreasing values of $\tau_{drag}$ (Figure \ref{fig:drag_zmz}). These test cases result in a transition climate at a $\tau_{drag}$ of 10 days and eyeball climates with $\tau_{drag}$ of 1 and 0.01 days, with a much larger day-night temperature contrast in the latter case (Figure \ref{fig:drag}). This result points to horizontal advection from the dayside to the nightside as a driver of the high-\pn climate transition.

\subsection{Water Vapour}

\begin{figure}[h!]
    \centering
    \includegraphics[width=\textwidth]{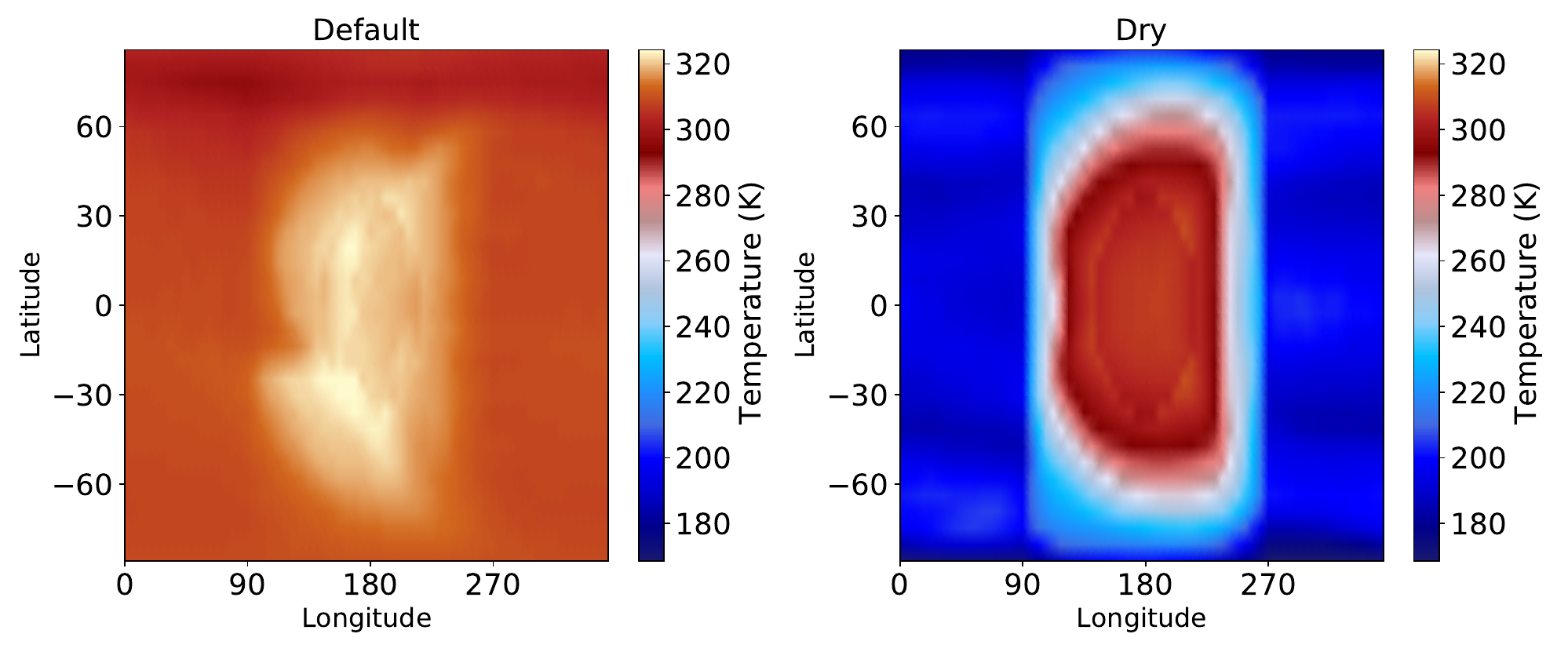}
    \caption{Surface temperature comparison between default and dry-atmosphere runs of a SubCont model with a \pn of 8 bars, instellation of 881.7~\wm, and 20\% dayside land cover. The default version is in the temperate nightside regime, whereas the dry version is in the eyeball regime with very cold nightside temperatures.}
    \label{fig:dry_comparison}
\end{figure}

To investigate the role of water vapour in heat transport, we run a simulation with surface evaporation turned off, such that there is water in the ocean, but it does not enter the atmosphere. We use the drag-free temperate nightside SubCont planet from Section \ref{sec:drag}. Figure \ref{fig:dry_comparison} compares surface temperature maps for the wet and dry atmosphere simulations. The dry version remains in the eyeball state, with a completely frozen nightside and a partially deglaciated dayside, including some ice-free ocean.

Water vapour plays a role in heat transport through evaporation from the dayside, advection to the nightside, and precipitation. It also contributes to the radiation balance by absorbing incident stellar radiation and outgoing longwave radiation emitted by the surface. These tests show that both advection and water vapour are required for the climate transition to the temperate nightside regime.

\subsection{Radiation Balance}

\begin{figure}[h!]
    \centering
    \includegraphics[width=\textwidth]{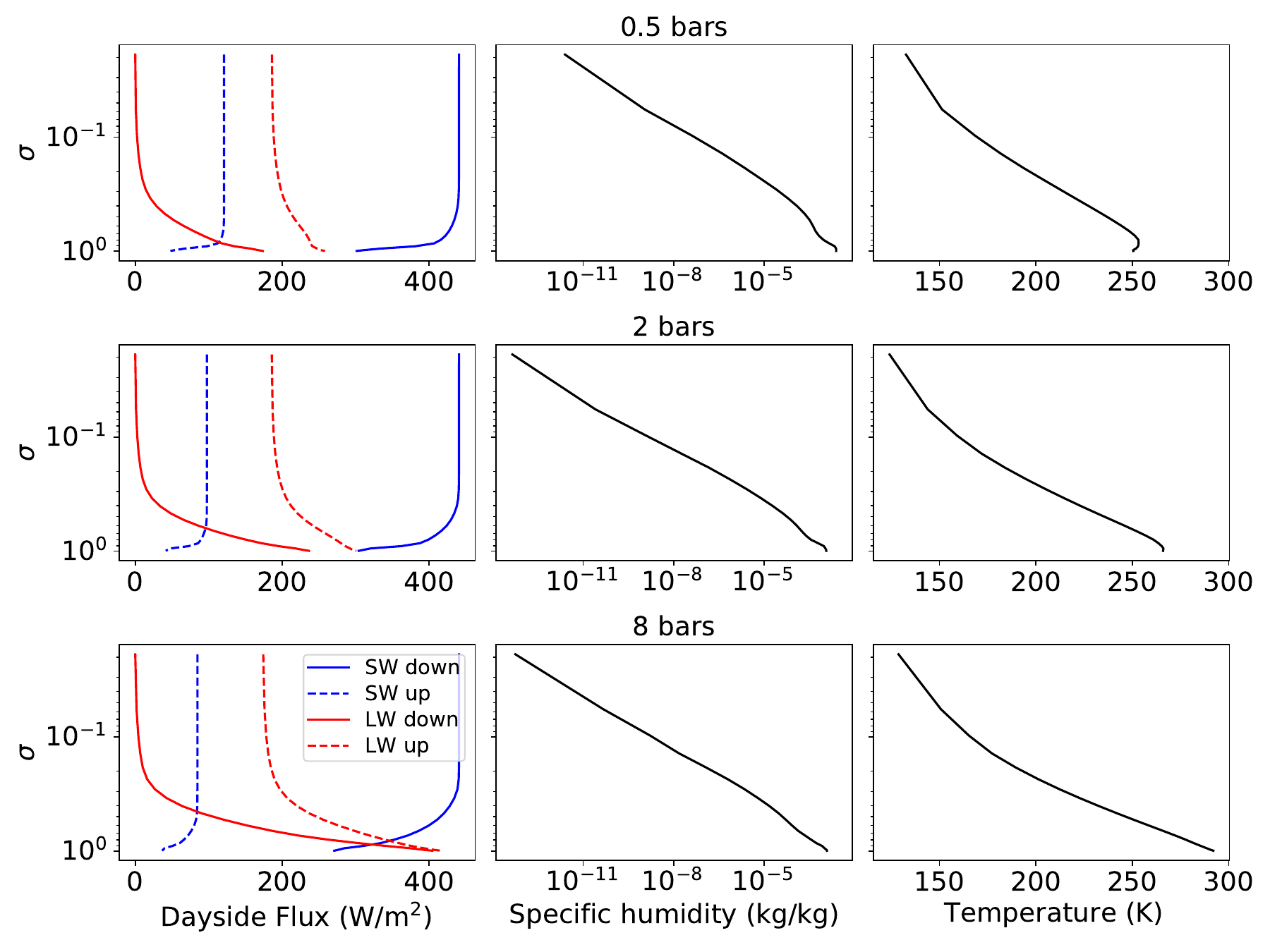}
    \caption{Dayside average fluxes, specific humidity, and temperature profiles (columns) for simulations with varying \pn (rows). The high-\pn planet is hotter and wetter. It also absorbs and releases energy higher in its atmosphere. Note that the y-axis is in $\sigma$ coordinates, where $\sigma=P/P_0$ for pressure level $P$ and surface pressure $P_0$, rather than total pressure.}
    \label{fig:transition_fluxes}
\end{figure}

We compare the shortwave and longwave fluxes at each vertical level to understand where in the atmosphere the energy is being absorbed (Figure \ref{fig:transition_fluxes}). Shortwave here corresponds to incident energy from the star, and longwave to energy emitted by the planet or its atmosphere. More shortwave energy is absorbed higher in the atmosphere as the pN$_2$ increases. The atmosphere is also more opaque to longwave radiation at high \pn. This stronger greenhouse effect results in a warmer surface.

\subsection{Stratospheric Humidity}

Hot planets with significant water vapour in their atmospheres are at risk of entering an uninhabitable moist greenhouse state, in which water vapour enters the stratosphere and can then be lost to space. \citet{Kasting1993} define a moist greenhouse limit as stratospheric humidity exceeding $3\times10^{-3}$~kg/kg, although this threshold can vary \citep{Wordsworth2014}. Our simulations do not have a stratosphere, so we instead look at the maximum value of specific humidity in the highest layer of the atmosphere. This value is on the order of $10^{-6}$~kg/kg for our wettest climates, and is several orders of magnitude lower in the majority of our simulations, so we conclude that these climates are not in a moist greenhouse state.

\subsection{Climate Cycling}\label{sec:pco2cycling}
To investigate whether there are multiple stable climate states for a given configuration, we gradually perturb the instellation or p\co of a simulation. We run these simulations at a resolution of T21 to allow them to reach an equilibrium state after each perturbations over several thousand years of total simulation time. We use a SubCont planet with 60\% dayside land cover, a pN$_2$ of 6 bars, and an instellation of 900~\wm. This simulation is initially in the transition regime. A similar planet with either a lower land fraction, a higher \pn, or a higher instellation would be in the temperate nightside regime, while one with a higher land fraction or a lower \pn or instellation would be in the eyeball regime. 

To test this planet's climate sensitivity and stability, we run an initial simulation with the above parameters for 400 years, which is longer than the typical time required to reach energy balance equilibrium, in order to obtain an initial stable climate state in the transition regime. We then increase the pCO$_2$ by 200~$\mu$bar or the instellation by 10~\wm and run the simulation for another 400 years, which results in a different stable climate state. We continue to increase the p\co in steps of 200~$\mu$bar or the instellation in steps of 10~\wm, running for 400 years at each step, until the planet reaches the temperate nightside regime, with no sea ice. We then decrease the p\co or instellation in the same manner until the ocean is completely ice-covered. We then reverse the sign of the perturbation again to return the climate to the temperate nightside state.

\begin{figure}[h!]
    \centering
    \includegraphics[width=\textwidth]{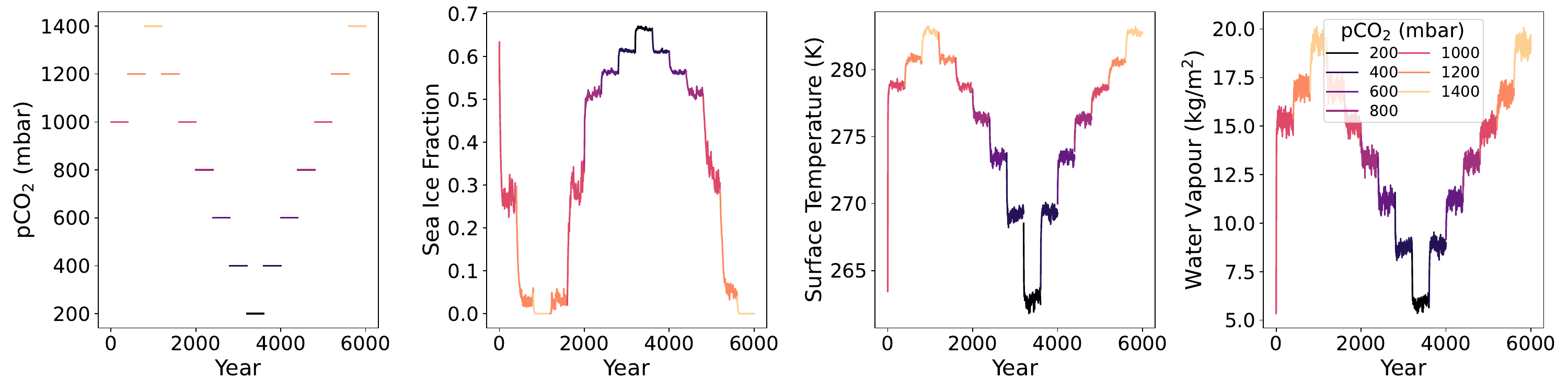}
    \caption{Left to right: p\co, sea ice fraction, globally averaged surface temperature, and globally averaged water vapour as a function of time for a SubCont planet with an instellation of 900~\wm and 60\% dayside land cover. The simulation is initialized with 1000~$\mu$bar of \co and run for 400 years, at which point its nightside is partly ice-covered. We increment the p\co in steps of 200 $\mu$bar per 400 simulation years. Line colours represent the p\co. There is no strong dependence of the climate state on initial conditions, but there is significantly more climate variability in the partially deglaciated states.}
    \label{fig:pco2cycle}
\end{figure}

Figure \ref{fig:pco2cycle} shows the sea ice fraction, average surface temperature, and atmosphere water vapour content for the p\co perturbation cycle described above. The ice cover is most sensitive to p\co around the nightside ice transition. We do not observe a hysteresis: for each p\co, the climate reaches the same final state from both directions. However, we note that there is more climate variability in the transition regime, when the nightside is partially deglaciated. This may be due to the way ExoPlaSim handles sea ice: each grid cell must be either fully ice-covered or fully ice-free \citep{Checlair2017}. Fluctuations in ice cover are therefore expected as ocean cells near the freezing point alternate between ice-covered and ice-free between timesteps.

\begin{figure}[h!]
    \centering
    \includegraphics[width=\textwidth]{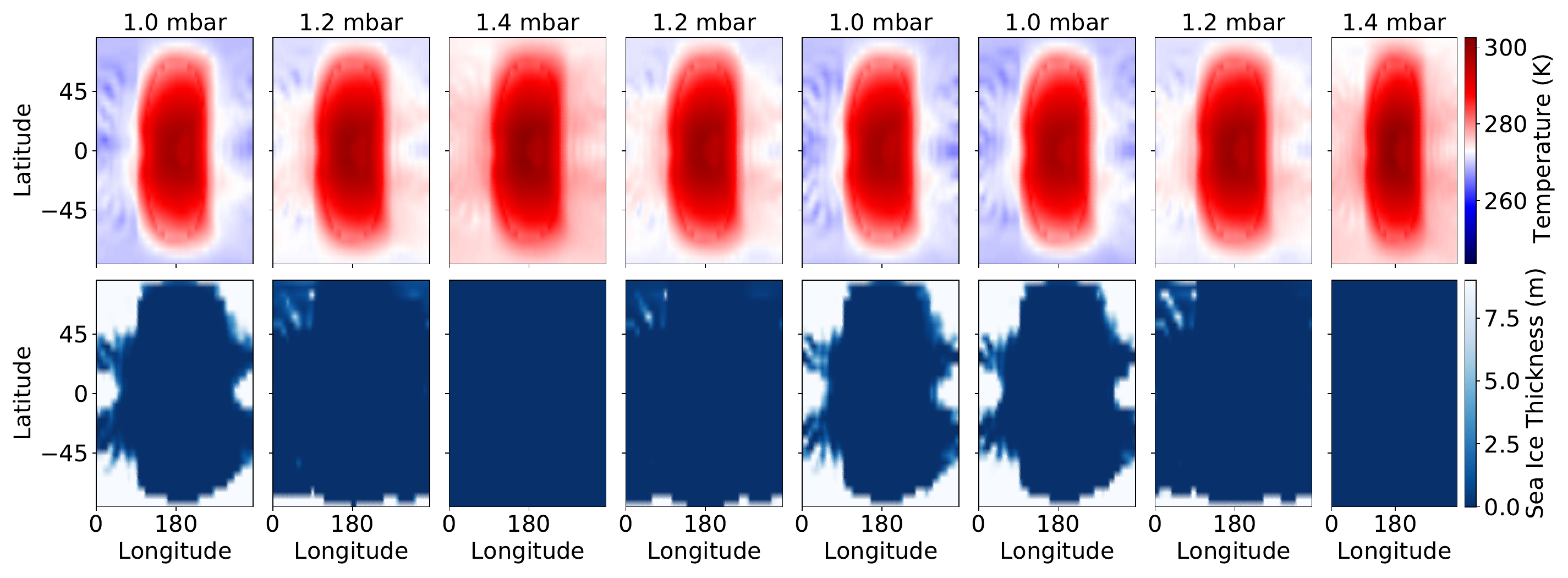}
    \caption{Surface temperature (top row) and sea ice thickness (bottom row) for selected simulations from Figure \ref{fig:pco2cycle}. Column titles are the p\co. Each is shown at the end of the 400$^{th}$ simulated year, right before the p\co is perturbed again. Panels are in chronological order from left to right, with some simulations omitted. At a given p\co, the climate always reaches a nearly identical equilibrium state, regardless of whether the preceding state was warmer or colder.}
    \label{fig:pco2cycle_maps}
\end{figure}

Figure \ref{fig:pco2cycle_maps} shows maps of equilibrium surface temperature and sea ice fraction for each cycle between the transition and temperate nightside regimes in Figure \ref{fig:pco2cycle}. We obtain one climate state per p\co regardless of the direction of the perturbation.

\begin{figure}[h!]
    \centering
    \includegraphics[width=\textwidth]{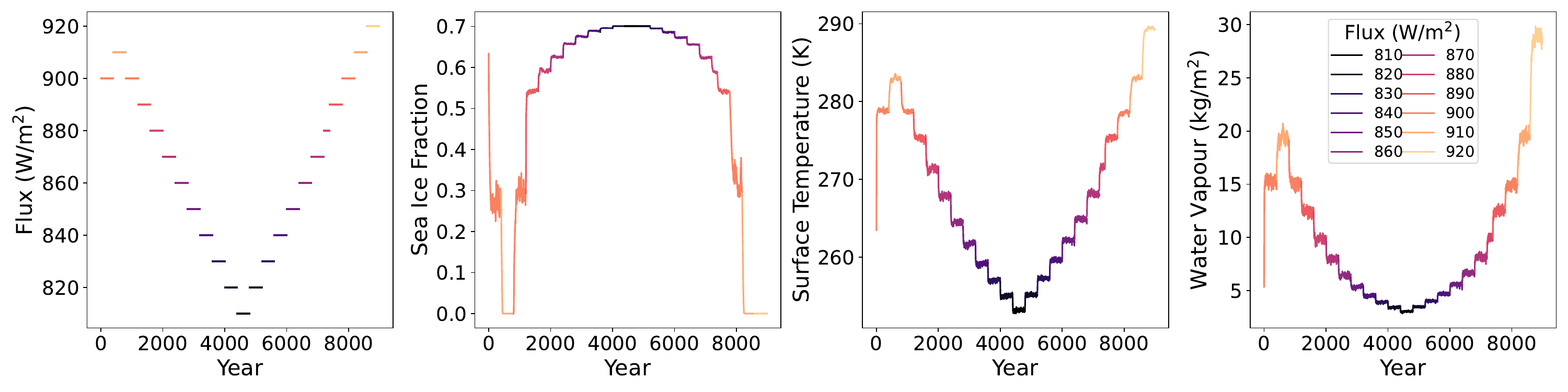}
    \caption{Left to right: incident flux, sea ice fraction, globally averaged surface temperature, and globally averaged water vapour as a function of time for a SubCont planet with 60\% dayside land cover, a constant p\co of 1000~$\mu$bar and varying instellation in steps of $\pm10$~\wm every 400 simulation years. Line colours represent the incident flux. Again, there is no strong dependence of the climate state on initial conditions, but there is significantly more climate variability in the partially deglaciated states.}
    \label{fig:fluxcycle}
\end{figure}

\begin{figure}[h!]
    \centering
    \includegraphics[width=\textwidth]{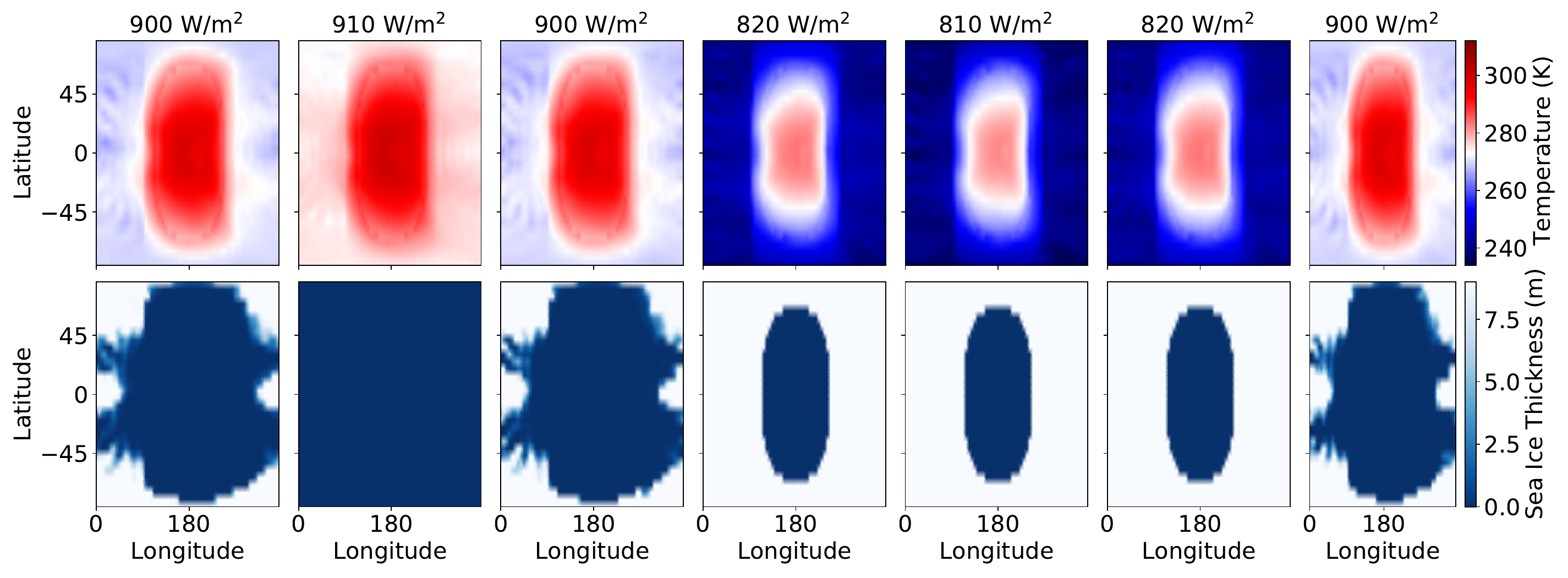}
    \caption{Surface temperature (top row) and sea ice thickness (bottom row) for selected simulations from Figure \ref{fig:fluxcycle}. Column titles are the instellation. Each is shown at the end of the 400$^{th}$ simulated year, right before the instellation is perturbed again. Panels are in chronological order from left to right, with some simulations omitted. At a given instellation, the climate always reaches a nearly identical equilibrium state, regardless of whether the preceding state was warmer or colder.}
    \label{fig:fluxcycle_maps}
\end{figure}

Figures \ref{fig:fluxcycle} and \ref{fig:fluxcycle_maps} show the climate trends and maps for the instellation perturbation experiment. Again, the climate cycles smoothly between states, with more instability in partially deglaciated states. The climate is most sensitive to changes in instellation around the transition regime: there is a difference of only 20~\wm between temperate nightside and eyeball states, whereas going from the transition regime to a fully ice-covered ocean requires further decreasing the instellation by 80~\wm.

Although these climate cycles are meant as a test of the robustness and stability of the climate states in these simulations, we note that a planet's instellation, surface pressure, greenhouse gas abundance, and water availability are all subject to changes over the course of its evolution. These experiments strengthen the conclusions of \citet{Checlair2017, Checlair2019}, who found that synchronously rotating planets cycle smoothly between snowball and eyeball states, without a snowball bifurcation. They found that the ice albedo feedback is weaker on synchronous rotators than on Earth because the permanent nightside does not receive any direct incident stellar radiation, so the albedo of nightside snow and ice is irrelevant. We further note that the albedo of snow and ice is much lower in the infrared, where M-dwarfs emit more of their energy, so the feedback is further weakened compared to planets with a Sun-like stellar spectrum.

\section{Impacts on Observations}\label{sec:obs}

At the time of writing, no atmospheres have yet been detected on M-Earths. Current and next-generation instruments will attempt to make such a detection over the coming years. In this section, we discuss how the large variety of climates presented in this paper would appear in transit and eclipse observations.

\subsection{Transmission Spectroscopy}

JWST will search for M-Earth atmospheres using transmission spectroscopy. These measurements are challenging because the signal from the planet's atmosphere is small compared to stellar noise and other uncertainty sources. As seen in Figure 6 of \citet{Macdonald2024}, the high-\pn, temperate nightside planets have the largest water vapour spectral features in synthetic transmission spectra, while low-\pn, high-land-fraction planets have much smaller transit signals.

\begin{figure}[h!]
    \centering
    \includegraphics[width=\textwidth]{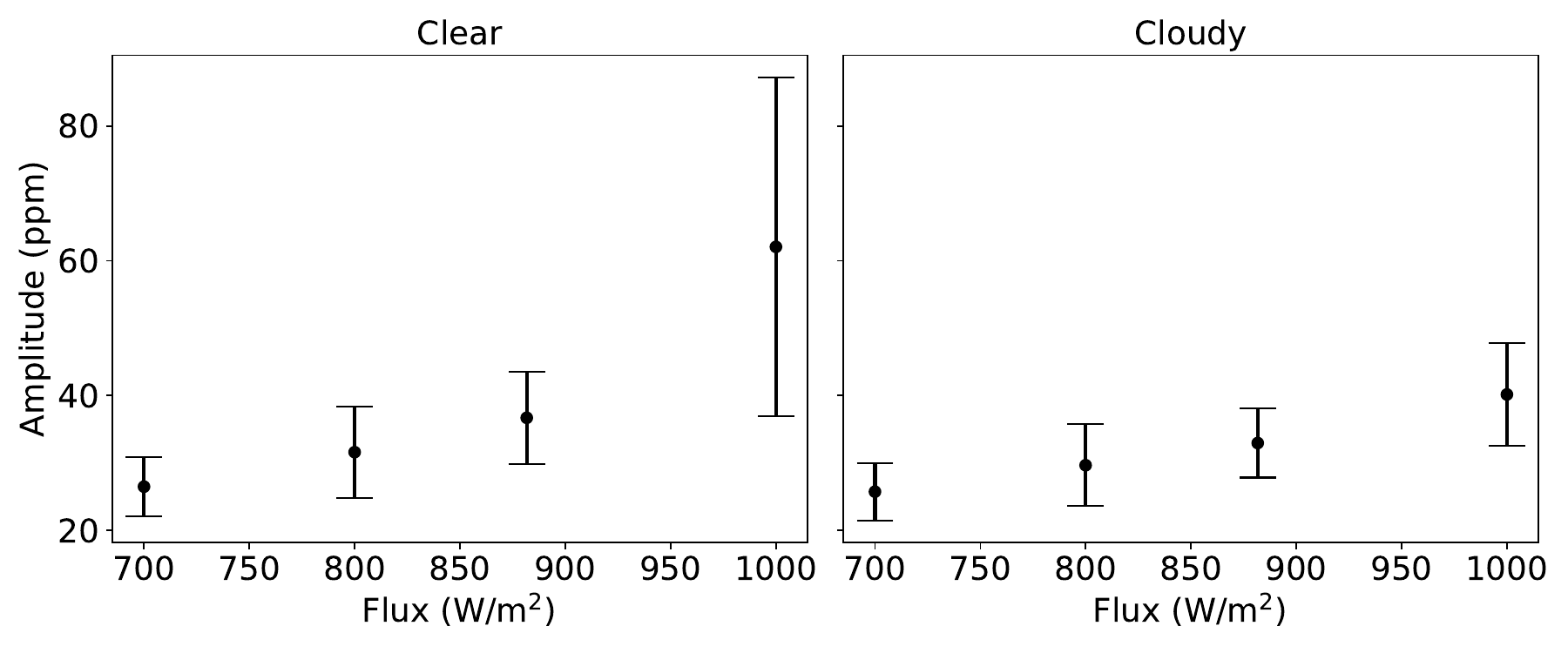}
    \caption{Water vapour transit amplitude in clear sky (left) and cloudy (right) synthetic spectra. Points are the average amplitude for all of the simulations of the corresponding flux, and error bars represent the standard deviation. Higher-instellation planets span a larger range of amplitudes, and are more affected by clouds, because these climates are warmer and wetter and have cloudier terminators.}
    \label{fig:transit_amplitude}
\end{figure}

This trend is even more apparent with the range of incident fluxes in the present study's simulations. Figure \ref{fig:transit_amplitude} shows the average water vapour amplitude as a function of incident flux for clear and cloudy conditions. As in \citet{Macdonald2024}, synthetic spectra are produced with petitRADTRANS \citep{Molliere2019}, with water vapour as the only absorber, and the amplitude is defined as the maximum differential transit depth of the $\sim$6 $\mu$m \water spectral feature. We find that a higher flux corresponds to both larger amplitudes and a larger spread, but that this spread is greatly reduced when clouds are included because they truncate the spectra higher up.

\bigskip
\bigskip
\bigskip
\bigskip
\bigskip
\bigskip

\subsection{Phase Curves}

It is also relevant to consider whether M-Earth climate regimes could be identified using thermal phase curves. These observations measure the planet's flux over a full orbital period, and therefore can provide temperature information. The phase curve amplitude is related to the day-night temperature contrast, which depends on the planet's climate regime.

Previous studies have used simple models of synthetic M-Earth phase curves for different climate states. \citet{Yang2013} found differences in phase curve amplitude and shape for slow rotators with different atmospheres and fluxes. They found that clouds cause large differences in hotspot offset for high-instellation cases. \citet{Haqq-Misra2018} found that phase curves could be used to differentiate between rotation-related climate regimes, with rapid rotators exhibiting smaller phase curve amplitudes and shifted hotspots. \citet{Komacek2019} compared theoretical bolometric phase curves calculated from the upwelling top-of-atmosphere longwave flux for M-Earths with a range of star temperatures and incident fluxes. They also found that high-instellation, rapidly rotating planets have much flatter phase curves and larger hotspot shifts than slower rotators with lower incident flux. 

We generate simplified synthetic phase curves for the simulations in this paper, assuming no orbital inclination. The thermal emission at a given orbital phase, $\phi$, is the integrated top-of-atmosphere (TOA) outgoing longwave radiation (OLR) in the direction of the observer. The TOA OLR of each grid cell in the direction of the observer, $f(\theta_i,\lambda_j)$ is the OLR multiplied by the cosine of the angle $\alpha_{ij}$ from the cell to the centre of the visible hemisphere, which is centred at longitude $\lambda=\phi$ and latitude $\theta=0$:
\begin{equation}
    f(\theta_i,\lambda_i,\phi) = OLR_{ij}\cos{\alpha(\theta_i,\lambda_j,\phi)}
\end{equation}
where
\begin{equation}
    \sin^2\alpha(\theta_i,\lambda_j,\phi) = \cos^2\theta\sin^2{(\lambda-\phi)} + \sin^2{\theta}.
\end{equation}
The flux, $F(\phi)$, is then the weighted average of $f$ over the surface area of the visible hemisphere:
\begin{equation}
    F(\phi) = \frac{\sum_{i,j}{f(\theta_i,\lambda_j},\phi)dA_{ij}}{\sum_{i,j}\cos\alpha_{ij}dA_{ij}}
\end{equation}
where $dA_{i,j}$ is the element of surface area for grid cell $(i,j)$.

\begin{figure}[h!]
    \centering
    \includegraphics[width=\textwidth]{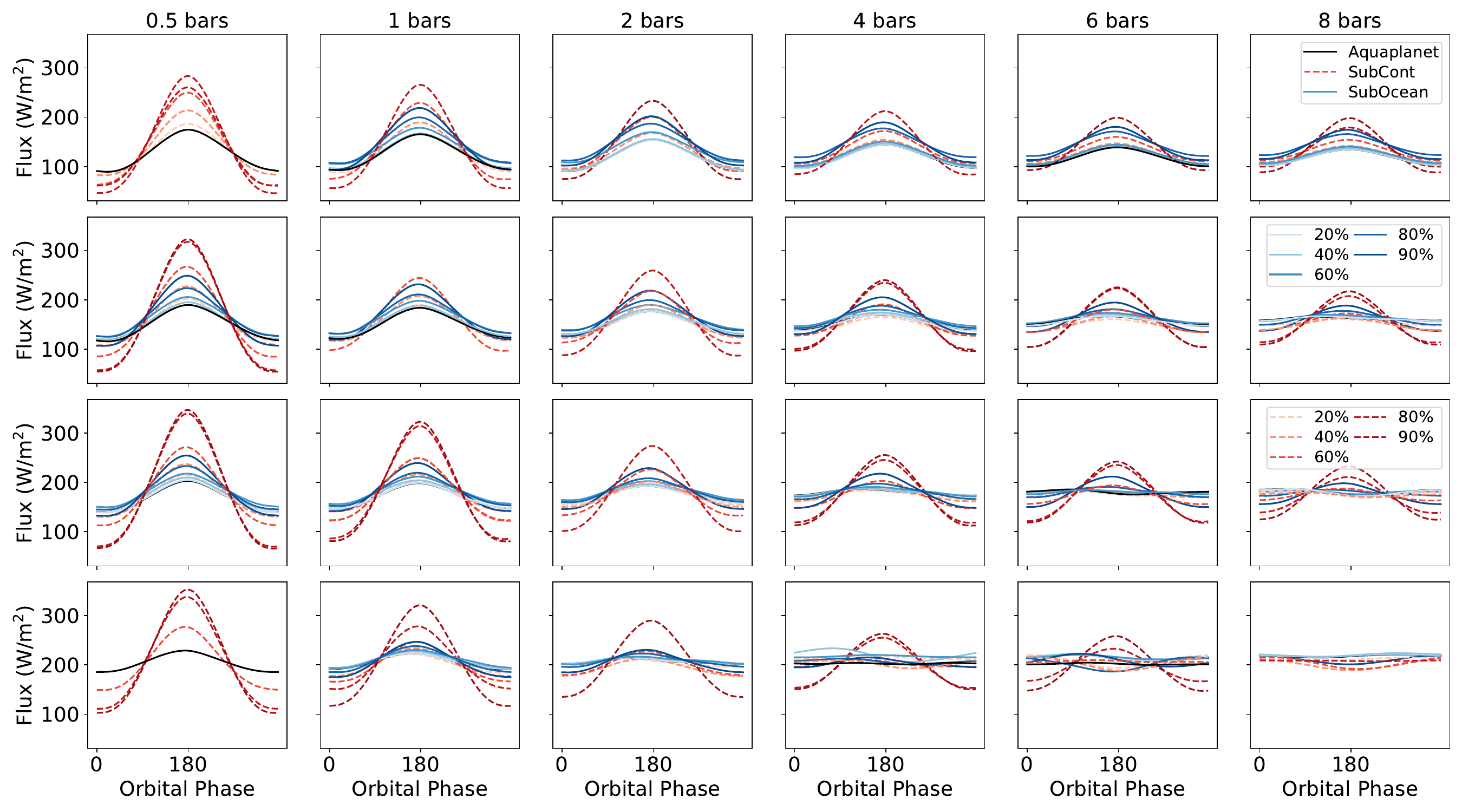}
    \caption{Theoretical phase curves for planets with varying \pn and land cover. Top to bottom: instellations of 700, 800, 881.7, and 1000~\wm. Legend percentages indicate dayside land fraction. Secondary eclipse (dayside emission) is at an orbital phase of $\phi=180^\circ$. Planets with higher \pn have flatter phase curves, especially at 1000~\wm where more of them are in the temperate nightside regime. However, the 700~\wm high-\pn snowballs also have small amplitudes, so knowledge of instellation and overall planetary flux will be needed to distinguish between temperate nightside and snowball climates.}
    \label{fig:phase_curves}
\end{figure}

Figure \ref{fig:phase_curves} shows synthetic thermal phase curves for our set of simulations. Amplitudes are largest for planets with low \pn in the eyeball regime. Low-instellation, high-\pn snowballs have low overall fluxes and small amplitudes, as their dayside and nightside temperatures are cold. Temperate nightside planets exhibit much smaller amplitudes and larger hotspot offsets, similar to the rapid rotators of \citet{Haqq-Misra2018} and high-instellation planets of \citet{Komacek2019}. However, rotation rate and instellation can be inferred from semi-major axis and stellar temperature for synchronous rotators, whereas \pn and land cover are unconstrained. Phase curve amplitude and shape may therefore provide more information about the latter two parameters.

\section{Discussion}\label{sec:discuss}

We have shown using ExoPlaSim simulations that at sufficiently high instellation and \pn, M-Earth climates transition from the eyeball regime with a frozen nightside to a ``temperate nightside" climate regime in which the entire planet is above freezing, the day-night temperature differences are small, and the horizontal wind speeds are greatly reduced. We also find a ``transition" climate regime between the eyeball and the temperate nightside, in which some nightside ice remains, particularly around the poles. The threshold \pn for a planet to transition at a given instellation is very sensitive to its land cover.

We have shown that this climate transition is driven by increased advection of heat to the nightside and heat transport by water vapour. We have found that the temperate nightside regime is stable, and shown that a planet can cycle back and forth between the eyeball and temperate nightside regimes in response to perturbations in instellation or p\co. We did not observe a snowball bifurcation, meaning that there is only one stable equilibrium climate state for a given combination of land cover, \pn, p\co, and instellation; the planet will reach this equilibrium state from warmer or colder initial conditions. This result is relevant because a planet's \pn, p\co, and instellation can vary over time. Earth has some nitrogen stored in its mantle, and the \pn of its atmosphere is thought to have varied over geological timescales \citep{Goldblatt2009}. A planet may therefore be susceptible to transitions between climate regimes over geological timescales.

However, the equilibrium state for a given configuration is very sensitive to water availability and thus land cover. Because this climate transition relies on the presence of water vapour in the atmosphere, we would not expect it to occur in the same way on planets with low water inventories. These are more likely to remain in eyeball states.

We have also shown that cold climates are more likely to be in the snowball state at low land fraction or high \pn due to the increased advection of heat to the nightside by more massive atmospheres. This means that the outer edge of the habitable zone could be at higher instellation for thinner atmospheres, and is sensitive to dayside land fraction.

\citet{Paradise2021} used a Sun-like spectrum to model Earth-like and synchronously rotating planets with varying instellation and \pn with PlaSim. They found that the transition between an ice-free and ice-covered dayside for synchronous rotators, or between an ice-free and ice-covered entire planet for Earth-like rotators, spans a smaller instellation range at high \pn. They attribute this effect to the decreasing surface temperature gradients with increasing \pn. Similarly, we find that high-\pn planets go from snowballs at 700~\wm to temperate nightsides at 1000~\wm, whereas lower-\pn planets are eyeballs through this entire instellation range.

\citet{Keles2018, Komacek2019, Paradise2021} found that for planets with Sun-like host stars, competition between Rayleigh scattering and pressure broadening causes a nonlinear temperature response to increasing \pn, with the warming effect of pressure broadening dominating at low surface pressures and the cooling effect of Rayleigh scattering dominating at higher surface pressures. \citet{Keles2018} further showed that above 4 bars, when Rayleigh scattering is neglected, increasing \pn also has a net warming effect of increasing magnitude. \citet{Paradise2021} showed that when Rayleigh scattering is turned off, surface temperature increases monotonically with \pn. Rayleigh scattering is much less important for the late M-dwarf host star used in our simulations, so the monotonic temperature increase we observe with increasing \pn is consistent with their results.

Our model atmospheres are composed of N$_2$ with trace \co and \water. We have not included other trace gases or attempted to model atmospheric chemistry or aerosols. \citet{Keles2018} found, using a 1D climate model coupled to a chemistry model, that changes in \pn affect the vertical profiles and can enhance spectral features of trace gases in an Earth-like atmosphere with a Sun-like stellar spectrum. However, some of these spectral features are then obscured by the strong effects of pressure broadening on \co and \water bands. A consistent chemistry model would be needed to understand these effects in M-Earth spectra. The results are also likely to be sensitive to the cloud parameterizations in both the GCM and the radiative transfer model.

Observations may eventually be able to differentiate between some of the more extreme climate states discussed in this paper. Temperate nightside climates exhibit small phase curve amplitudes and prominent transit water vapour features, attributed to their minimal temperature gradients and high water vapour content. In contrast, thin atmospheres and dry surfaces are associated with much larger phase curve amplitudes and significantly smaller water vapour transit signals. Our high-instellation simulations demonstrate the greatest variability in transit amplitude, a variability that is greatly reduced when clouds are included in the radiative transfer calculation.

The model planet used in this paper is small, to optimize its atmosphere for transmission spectroscopy. Phase curves would be easier to observe on a larger planet. Larger day-night temperature contrasts would also be expected in that case.

A mission sensitive to longer wavelengths, such as MIRECLE \citep{Mandell2022MIRECLE}, would be able to detect and characterize M-Earth atmospheres using planetary infrared excess \citep{Stevenson2020PIE}. In this method, the planet and star are observed simultaneously at a wide range of infrared wavelengths, and both their fluxes are fitted to the resulting spectrum. The planet's signal can then be isolated. The planet's signal is stronger in the thermal infrared because it is cooler and therefore peaks at longer wavelengths than the star.

\section*{Data Availability}
The simulations in this study and files needed to reproduce them are available in Borealis repositories \citep{Macdonald2024data, Macdonald2024pN2data}.

\begin{acknowledgments}
EM is supported by a Natural Science \& Engineering Research Council (NSERC) Post-Graduate Scholarship and by the University of Toronto Department of Physics. KM is supported by NSERC. CL is supported by the Department of Physics. The University of Toronto, where most of this work was performed, is situated on the traditional land of the Huron-Wendat, the Seneca, and the Mississaugas of the Credit. This study made substantial use of supercomputing resources; although most of the energy used comes from low-carbon sources, we acknowledge that building and operating these facilities has a significant environmental impact.
\end{acknowledgments}

\software{ExoPlaSim \citep{Paradise2022}, petitRADTRANS \citep{Molliere2019}}

\bibliography{references}{}
\bibliographystyle{aasjournal}

\end{document}